\documentstyle[11pt]{article}
\begin{document}
\newtheorem{thm}{Theorem}[section]
\newtheorem{defin}[thm]{Definition}
\newtheorem{lemma}[thm]{Lemma}
\newtheorem{propo}[thm]{Proposition}
\newtheorem{cor}[thm]{Corollary}
\newtheorem{conj}[thm]{Conjecture}

\centerline{\Large \bf ON ALGEBRAIC STRUCTURES IMPLICIT IN}
\centerline{\Large \bf TOPOLOGICAL QUANTUM FIELD
THEORIES}

\bigskip

\centerline{\parbox{4in}{Louis Crane and David Yetter \\ Department of
Mathematics \\ Kansas State University \\ Manhattan, KS 66506}}

\bigskip

\section{INTRODUCTION}

\bigskip

In the course of the development of our understanding of topological
quantum field theory (TQFT) [1,2], it has emerged that the structures
of generators and relations for the construction of low dimensional
TQFTs by various combinatorial methods are equivalent to the
structures of various fundamental objects in abstract algebra.

Thus, 2D-TQFTs can be constructed from commutative Frobenius algebras [3]
or from semisimple associative algebras [4]; while 3D theories can be
constructed either from nicely behaved braided monoidal categories
[5,6 7,8,9]
or from Hopf algebras [10,11].

In [12], a possible method was proposed to extend this picture to D=4.
Namely, it was shown how to construct a 4D TQFT from a new type of
algebraic structure called a Hopf category.

The purpose of this paper is to show that
under physically reasonable hypotheses, the seemingly exotic algebraic
structures used in the constructions above arise naturally from
3D and 4D TQFT's.  We shall show that any 3D-TQFT with a property which we
call
factorizability, which
any TQFT which came from a path integral with a
topological lagrangian would be expected to satisfy, contains a
braided monoidal category Hopf in its structure, and that this category
arises from a generalized Hopf algebra by a construction first proposed
by Yetter. We shall show moreover that any factorizable 4D-TQFT gives
rise to a Hopf category object in a certain reasonably concrete bicategory.
This theorem lends weight to the conjecture in [12] that the
4D-TQFT which is believed to be constructable from Donaldson-Floer
theory is related
to the Hopf category constructed in [12] from the canonical basis of a
quantum group [13].

The importance of the procedure we outline in this paper is greatly
increased by the recent breakthrough in the understanding of
Donaldson-Floer theory made by Witten [14]. The pair of differential
equations whose solutions are related to DF theory in this new
approach is much more tractible than the self duality equation [15]. In
particular, there is a well behaved version of them on manifolds with
boundary. Thus, we can take the geometric constructions in this paper
as a prescription: consider the space of solutions of Witten's equations on the
manifolds with boundaries or corners we are examining, there must then
appear certain algebraic operations on them, from which the TQFT can
be reconstructed.

The contents of this paper are as follows: Section \ref{fact3} gives the
definition of a 3D-TQFT with factorizability and physical motivation
for the definition. In Section \ref{hopfalg},
we prove
that every 3D-TQFT with factorizability contains a Hopf algebra
object, and show the relation between this object and the category
used to define factorization. Section \ref{hopfcat} recapitulates the
definition of a Hopf category. In
Section \ref{fact4}, we explain the extension of the definition of
factorizable
TQFT to D=4, and prove of the main theorem in 4D. Finally, in
Section \ref{concl}, we outline some extensions of our argument and suggest
directions for further work.

Throughout the paper all manifolds (with or without boundary or corners)
are compact and oriented.

\section{ FACTORIZABLE 3D-TQFT \label{fact3}}

The subject of TQFT began with the study of path integrals for
lagrangians with topological invariance. A rigorous treatment of this
approach is not within the reach of the mathematics of our time.
Nevertheless, it is possible to make formal manipulations of path
integrals to deduce that the theories derived from them should have
certain properties. We define a TQFT with these properties as
factorizable.

The properties come from two aspects of the theory of path integrals.
One is the idea that since a path integral is a ``sort of'' integral for
each point in a space, we can  separate it into integrals over parts
of a space, by a ``sort of'' Fubini's theorem. The other is that the
topological lagrangians possess a large gauge symmetry, with respect
to which physical states must be invariant. If we cut space up along
submanifolds of codimensions one and two, we get states with boundary attached
to codimension one submanifolds with codimension two boundaries which
transform non-trivially under the quantum version of the gauge
symmetry on the codimension two submanifolds.. It is this gauge symmetry
which is responsible for the
appearance of tensor categories or Hopf algebras in the structure of a
TQFT.

A formal derivation from a path integral would serve no mathematical
purpose, since path integrals themselves are not rigorously defined.
Let us, then
simply write down a set of axioms for a factorizable 3D-TQFT. The
argument from the path integral is given in the special case of CSW
theory in [14].

Let us begin by recalling the structure of a cobordism category. The
category of oriented n-dimensional cobordisms has oriented compact n-1
dimensional manifolds as objects and cobordisms as morphisms. (A
cobordism from M to N is an oriented n-dimensional manifold P with boundary,
together with an oriented diffeomorphism between the boundary of P and
$M^* \bigcup N$.)
Composition of morphisms comes from gluing of manifolds along shared
boundary components.

The category of oriented n-cobordisms has the natural structure of a
tensor category with duality. The tensor product is direct sum and the
duality is reversal of orientation.

The most elegant definition of an n-dimensional TQFT is that it is a
monoidal
functor from the category of oriented n-cobordisms with disjoint union
as tensor product to the category
{\bf VECT} of
finite dimensional vector spaces with the usual tensor product (i.e. a
functor which preserves tensor product up to canonical coherent
isomorphism). It is a point often missed that this suffices---that
a manifold with opposite orientation is sent to the dual space of the
image with the given orientation is an easy theorem, not a necessary
part of the definition.

We often modify the definition of TQFTs by modifying the cobordism
category. For instance, we can specify a framing of the tangent
bundle of the cobordisms and of a formal neighborhood of the closed
manifolds.  Another possibility is to include insertions of submanifolds
 in the
manifolds and matching insertions in the cobordisms. We also refer to
tensor and duality preserving functors from such modified cobordism
categories to {\bf VECT} as TQFTs.

Let us spell out this definition for the less categorically inclined.
An n-dimensional
TQFT assigns a vector space to each oriented n-dimensional manifold,
and a
linear map to each oriented cobordism in such a way that the
composition of cobordisms corresponds to the composition of linear
maps, the disjoint union of manifolds gets the tensor product of
vector spaces, and the manifold with opposite orientation gets
assigned the
dual space.

Thus, for n=3, a TQFT assigns a vector space to a surface, and a
linear map to a 3-dimensional cobordism.

Let us note that since the boundary of a cobordism is a disjoint union
of two manifolds, one with reversed orientation, it is equivalent to
assign a linear map to a cobordism, or a vector in the vector space on
the boundary to a manifold with boundary. It follows from this observation
that the invariant of closed 3-manifolds arising from a TQFT con be
viewed as the dual pairing of vectors associated to 3-manifolds
with (common, but oppositely oriented) boundary. This is, of course,
Atiyah's original view.

A 3D-TQFT with factorizability has an analogous structure one layer
farther down in dimension so that we can cut surfaces along sets of
circles and write the vector space on the surface as the hom-space
between objects associated to the pieces (the categorical analogue of
a dual pairing!).

To set up the formal definition embodying this notion, we must
remind the reader that a finitely generated
 semisimple linear category is one in which each
object is isomorphic to a direct sum of irreducible (simple) objects chosen
from a finite set of such objects, hom-sets are complex
vector spaces, and composition is bilinear. As categories, they are equivalent
to ${\bf VECT}^n$ for some $n$. For the theory of such
categories, also called {\bf VECT} modules, see [16]. As shown in [16],
these categories form a monoidal bicategory:
\footnote{Kapranov and Voevodsky call the structure a (monoidal) weak
2-category.  Their notion of a weak 2-category has long been
studied by categorists under the name ``bicategory'' (cf. Benabou [17]).
We here adopt the older name.  For terminology in this regard, and
details of some of the more abstact points of category theory, we refer
the reader to a recent paper of Gordon, Power and Street [18] and
references found therein.} objects are {\bf VECT} modules,
1-arrows are exact {\bf C}-bilinear functors, 2-arrows are natural
transformations, and the tensor product is given up to canonoical
equivalence by using pairs of the generating simple objects in the
tensorands as a set of generating simple objects.

Similarly observe that there is a monoidal bicategory of 3-dimensional
cobordisms with corners, $3-cobord_2$: its objects are 1-manifolds, its
1-arrows are (2-dimensional) cobordisms of 1-manifolds, and its
2-arrows are cobordisms with corners between pairs of 2-dimensional cobordisms
with the same source and target.
To be precise, a 3-dimensional cobordism with corners is a 3-manifold
with corners, whose boundary is a union along a family of circles
 joining corresponding boundary components
of the two surfaces. The 1-dimensional composition of 1-arrows and
2-dimensional composition of 2-arrows are just given by glueing target
to source. The 1-dimensional composition of 2-arrows consists of glueing
along the corners and glueing on a ``collar'' as shown in Figure 1.
It is trivial to verify that disjoint union gives
this bicategory the structure of a monoidal bicategory.

In what follows, we shall refer to a cobordism (resp. cobordism with
corners) as ``trivial'' if its underlying space
 is a product of one of its boundaries with the interval (resp. a product
of one of its boundary strata with the interval modulo collapsing
the product of the bounding corner with the interval back onto the corner).
Note that a trivial cobordism or cobordism with corners need not be
the identity cobordism---the attaching maps at the
ends could be different. However, a trivial cobordism is manifestly
invertible.

In Definitions \ref{3dfact} and \ref{4dfact} below, the non-categorically
minded reader is advised on first reading to read only the
bold-faced portions of the definitions. These give the essential flavor
of the definition, without going into excessive categorical detail.

\begin{defin} \label{3dfact}
A {\bf 3D TQFT with factorization} is a monoidal bifunctor
from $3-cobord_2$ to ${\bf VECT}-mod$.

Less briefly, but more intelligibly
to the non-categorically minded, this {\bf entails an assignment of}

\begin{enumerate}
\item  {\bf A finitely generated
semisimple {\bf C}-linear category to each compact 1-manifold.}
\item An exact {\bf C}-linear functor to each 2-dimensional cobordism. In
particular, since an exact
{\bf C}-linear functor from {\bf VECT} to a semisimple
{\bf C}-linear category is completely determined by the image of {\bf C},
we have {\bf a choice of an object in the category associated to the boundary
of each oriented surface with boundary, and more particularly, we have
an assignment of a vector space to every closed oriented surface.}
\item A natural tranformation to each 3-dimensional cobordism with
corners.  In particular {\bf for a 3-manifold with boundary and corners
consisting of two surfaces with boundary sharing their common
boundary as a corner, we have a map in the
category associated to the boundary of the surfaces between the
objects associated to the surfaces.}  Likewise, since the empty surface
is assigned {\bf C}, {\bf a 3-manifold with boundary is assigned a vector
in the vector space associated to its boundary, and finally
a 3-manifold without boundary is assigned a number.}
\end{enumerate}

Moreover, these assignments will satisfy:

{\bf
\begin{enumerate}
\item  The disjoint union of two 1-manifolds gets the tensor product in the
sense of [16] of
the semisimple categories attached to the parts. The empty 1-manifold
will be assigned {\bf VECT}.
\item The 1-manifold with opposite orientation is assigned the dual
category. (cf. Yetter [19])
\item The disjoint union of surfaces is assigned the tensor product of
the vector spaces on the surfaces.
\item The surface with opposite orientation is assigned the dual vector
space.
\item If we cut an oriented surface along a 1-manifold (union of circles),
the vector space on the closed surface is naturally isomorphic to the
hom set of the two objects in the category corresponding to the cuts
which correspond to the two surfaces with boundary. A similar result
holds for the case when we cut a surface with boundary and take hom
with respect to the ``tensor indices'' corresponding to the cuts only.
\item If we cut a 3-manifold along a surface with boundary, the number
invariant of the manifold is the dual pairing of the vectors
associated to the two manifolds with boundary.
\item If we join two cobordisms with corners to form a cobordism, the
linear map associated to the cobordism is the hom of the two linear
maps. If we join two cobordisms with corners along a surface with
boundary to form a new cobordism with the same corner, the map
corresponding to the new cobordism is the composite of the old maps.
\end{enumerate}
}
\end{defin}

The reader will no doubt have noticed that these assignments and
conditions fall into two analogous tiers, with semisimple linear
categories closely paralleling vector spaces. The situation for D=4
will be closely analogous again, with a third categorical tier.

It follows from these axioms that the category on a circle has an
associative tensor product, corresponding to the three holed sphere,
or trinion, with associativity constraints given by trivial cobordisms
with corners. Moreover, the category must be braided, again with structure
maps given by trivial cobordisms with corners.
It is the careful working out of analogous arguments one
dimension up which gives rise to the Hopf algebra object.

This definition is similar to ones proposed by Kazhdan [20], Walker [8] and
Lawrence [21], except
that they were not motivated by the gauge group of a lagrangian.

It is straightforward to generalize this definition to various
modifications of the cobordism categories as mentioned above. Such an
augmented TQFT will also be called factorizable.

\section{ THE HOPF ALGEBRA OBJECT \label{hopfalg}}

Given a factorizable 3D-TQFT, ${\cal T}$,  we obtain an object in the
category on the circle (a tensor category, as we mentioned above),
corresponding to the once punctured torus. Let us denote this object
as $O_ {\cal T}$.

\begin{thm} \label{firstthm} For any factorizable 3D-TQFT ${\cal T}$,
the object $O_
{\cal T}$ admits a natural structure as a Hopf algebra object in the
category on the circle. There is a natural isomorphism with the dual
object, which is also a Hopf algebra object.
\end{thm}

\noindent{\bf Proof:} The proof is essentially pictorial. In each case we
show a cobordism with corners (or composition of such) with a neighborhood
of the collar removed for clarity:  in all cases the bottom end of
the drawing should have the annulus shrunk to a circle.
The Hopf algebra object is the image of the torus with a disk removed
portrayed in Figure 2. Its tensor with itself as an object in the
category on the circle is given by Figure 3. The cobordisms (with
corners) which give
the product, unit, coproduct, and counit are shown in Figure 4.
Figure 5 shows the
antipode. Figures 6 and 7 give the proof of associativity, observe that
the inner and outer shells of the two figures are identical with respect
to the markings (which give the attaching map for the boundary), as is the
space between, only the division into composed cobordisms differs.
The proof of coassociativity is obtained by turning Figures 6 and 7 inside-out.
The proof of (half of) the antipode axiom is given in Figure 8 (the
other half being exactly similar). The reader should note that the
inner-most shell in the bottom portion of Figure 8 is not linked with the
hole in the outer shell, and that tracing the stripe shows that the
attaching map is the same as in the upper portion of Figure 8. The proofs
of the unit and counit axioms can be readily drawn by the reader.  The most
difficult axiom to verify is, as usual the connecting axiom. Observe that for
an object in a braided monoidal category, the connecting axiom must be taken
in the sense of Majid's braided Hopf algebras [22].  The one side of the
connecting axioms (multiply then comultiply) is shown in Figure 9.
Figures 10 through 12 give the maps which are composed to swap factors in the
comultiply then multiply side of the axiom, while Figure 13 gives that
side of the axiom (with the space between the second and third shells
reading inward given by the composition of Figures 10 though 12.
Again careful perusal shows that the space between the outer-most and
inner-most shells in Figures 9 and 13 is identical, as are the markings
indicating the attaching maps.

The isomorphism with the dual is depicted in Figure 14.  $\Box$
\smallskip

We note
that the quantum group associated with a Borel subalgebra is
isomorphic to its dual.

It is tempting now, to identify the category associated to the circle
with the representations (say modules) of the Hopf algebra.  This however
would be a mistake.  First, there is not necessarily an ``underlying
vector-space'' or fibre functor from the category to {\bf VECT}.  Even
were there such a functor, however, this would be a mistake.  In fact,
the relationship between the category and the object is more subtle.
To state it we need to recall one of the variants of Yetter's notion
of crossed bimodule [23] given by Radford and Towber [24].
\footnote{In Radford
and Towber's terminology [24] what was defined in Yetter [23] were ``left-left
Yetter-Drinfel'd structures'', while what we use here are ``left-right
Yetter-Drinfel'd structures.''}

\begin{defin}
A {\bf left-right crossed bimodule} over a Hopf-algebra $A$ is a vector-space
equipped with a left $A$-comodule structure and a right $A$-module structure,
moreover satisfying the condition (in modified Sweedler notation):

\[ \Sigma a_{(1)}\cdot m_{<1>} \otimes a_{(2)}\cdot m_{(2)} =
\Sigma (a_{(2)}\cdot m)_{<1>} \otimes (a_{(2)}\cdot m)_{(2)}a_{(1)} \]

\end{defin}

	Now, in terms of commutative diagram, this is given in Figure 15.
Observe that in two places, the symmetry map occurs.  Unfortunately, the
category associated to the circle in a factorizable 3D-TQFT is only braided,
so there is a question what the correct generalization is.  It turns out
that it is given by

\begin{defin}
A {\bf left-right crossed bimodule} over a Hopf-algebra object $H$ in a braided
tensor category is an object equipped with both a left $H$-comodule
structure and a right $H$-module structure such that the diagram in
Figure 16 commutes. A map of left-right crossed bimodules is a map
in the ambient category between crossed bimodules which is both a left
comodule map and a right module map.
\end{defin}

	The reader should note that in both Figures 15 and 16 we have
used the coherence theorem of Mac Lane [25] to suppress mention of the
associativity natural transformation.

	We can now show the following:

\begin{thm} \label{secondthm} Let ${\cal T}$ be a factorizable 3D-TQFT and let
 $\cal C$ be the category associated with the circle, and
$H$ be the Hopf-algebra object in $\cal C$ associated to the torus
with a disk removed, then every object associated to a
surface with a single boundary component
is a left-right crossed bimodule over $H$, and the map
between objects associated to any cobordism with a single circle
as corner is a crossed bimodule map. Moreover, the finite set of generating
objects used to define factorizability may be taken without loss of generality
to be left-right crossed bimodules.
\end{thm}

\noindent{\bf Proof:} Except for the last statement the proof is again
pictorial.  Figure 17 (resp. 18) shows the action (resp. coaction).
That the action is associative in the appropriate sense is verified in
Figure 19. Verification of coassociativity, unitalness and counitalness
of the action are similar and left to the reader. The verification of
the left-right crossed bimodule axiom is given in Figures 20 through 27 .
Figures 20 and 21 are composed to give Figure 22, giving the top
way around the diagram in Figure 16.  Figures 23, 24, 25 and the two
parts of Figure 26 are composed to give Figure 27 (in
Figure 26, the curves labelled 1 in each part correspond), giving the bottom
way around the diagram in Figure 16.

The remaining statement is a matter of algebra.  First, notice that we may
discard from the description of a factorizable 3D-TQFT any of the simple
generating objects that do not occur as direct summands of any object
associated to a surface with a single boundary component.  Now, observe that
each simple object is a summand of the object associated to a surface with
as single boundary component inherits a comodule structure.  In general
it would not inherit a module structure. In this case, however, the
Hopf-algebra object is self-dual, so a module structure is a comodule
structure for the dual Hopf-algebra object, and is thus inherited also.
Of course the restrictions necessarily satisfy the same compatibility
condition as the original action and coaction did, so the subobject is
a crossed bimodule. $\Box$

\section{ HOPF CATEGORIES \label{hopfcat}}

The passage from D=3 to D=4 will have the effect of lifting our
constructions by one categorical level. The analog of Theorem \ref{firstthm}
 will
accordingly produce a Hopf category. The notion of Hopf category was
introduced in [12]. For completeness, we repeat the definition here.

Let us note that there actually exist Hopf categories associated to
the quantum groups. This highly nontrivial fact follows from work of
Lustig [13].

\subsection{ Categories and Algebras}

We are now going to construct an analog of  the structure of a Hopf
algebra on a tensor category of a very special type.

To explain the idea of an analog of an algebraic structure on a
category, let us think briefly about the category {\bf VECT} of finite
dimensional vector spaces.
This category possesses two product operations, $\oplus$ and $\otimes$
and special objects {\bf 0} and {\bf 1}
with the properties
\smallskip

$(A\otimes B) \otimes C \cong A\otimes (B\otimes C)$

$(A \oplus B)\oplus C \cong A\oplus (B \oplus C)$

$A\otimes (B \oplus C) \cong (A\otimes B) \oplus (A\otimes
C)$

$A \oplus {\bf O} \cong A$

$A \otimes {\bf 1}\cong A$

$A \oplus B \cong B \oplus A$
\smallskip

These isomorphisms satisfy certain equations, called coherence relations.
(cf. Laplaza [26])
This is completely parallel to the definition of a ring.
 We describe this by saying that
{\bf VECT} is a {\bf ring category.} This structure is a
categorical analog of a ring.

	Other ring categories include various categories of (differential)
graded vector-spaces or modules.

What has happened is that equations in the ring correspond to
isomorphisms in the category. There are then natural equations
that the isomorphisms should satisfy, so that combining them in
different orders to produce a larger isomorphism always gives
consistent results. These were termed ``coherence relations''
by MacLane [25]. The coherence relations corresponding to the
commutative and associative laws are the Stasheff pentagons and
hexagons (cf. [16]).

Thus, if we replace an algebraic structure by a categorical analog, its
axioms will hold only up to natural isomorphisms, which in
turn must satisfy a new set of more complex equations, which are
its coherence relations. One of the fundamental ideas of the
dimensional ladder is that if we start with an algebraic structure
which can be used to construct a TQFT, than the coherence relations of
a categorical analog of it are just right to construct a TQFT in one
higher dimension. Our use of a Hopf category in 4D-TQFT is an
application of this idea.

\subsection{ Hopf Categories}

Now let us describe the structure of a Hopf category.

\begin{defin} A category is semisimple if each object is a direct
sum of simple objects (objects with no nontrivial sub- or quotient
objects). A semisimple category is finitely generated if it has only
finitely many inequivalent irreducible objects. In this paper, we will
vvvonly consider finitely generated categories (in order to make all
sums finite).
\end{defin}

\begin{defin} A category is linear if the set of morphisms has the
structure of a vector space, and composition is bilinear.
\end{defin}

\begin{defin} If {\bf R} is a ring category, then {\bf M} is a
{\bf module category} over {\bf R} if {\bf M} has an associative direct sum
and we are given a functor ${\bf R \times M
\rightarrow M}$ (denoted as multiplication) such that

\smallskip

$A_1 \otimes ( A_2 \otimes R) \cong (A_1 \otimes A_2)
\otimes R$

\smallskip

$(A_1 \oplus A_2) \otimes R \cong (A_1 \otimes R) \oplus (A_2
\otimes R)$

\smallskip

$A \otimes (R_1 \oplus R_2) \cong (A \otimes R_1) \oplus
(A \otimes R_2)$

and the isomorphisms satisfy the natural coherence relations
(cf. [16]).
\end{defin}

The concept of module category is the categorical analog of the
concept of a module.

\begin{defin} An {\bf algebra category} is a a ring category which is also
a {\bf VECT} module such that $\oplus$ is a  module map and $\otimes$ is
a module map in each variable separately.
\end{defin}

Note that a {\bf VECT} module must be linear.

{\bf VECT} modules are a categorical analog of vector spaces, so
algebra categories are categorical analogs of algebras.

Now recall that the dual ${\cal C}^op$ of a category $\cal C$ has the same
objects as the
category, but with morphisms reversed. Similarly, the dual of any
algebraic construction has diagrams corresponding to the first one,
but with arrows reversed. The dual of a {\bf VECT} module has a
natural
structure as a {\bf VECT} module, including a natural direct sum.
Less familiar is the fact (cf. Yetter [19]) that in the case of
{\bf VECT} modules, the dual category $\cal C$ is equivalent to the
hom-category $[{\cal C},{\bf VECT}]$, and thus gives rise to
a {\em contravariant} functor from {\bf VECT} modules to {\bf VECT}
modules.

\begin{defin} A coalgebra category is a {\bf VECT} module category
whose dual is an algebra category when dual category is understood
in the contravariant sense.
\end{defin}

This is equivalent to a {\bf VECT} module category with a
comultiplication functor
 $\Delta :{\bf A}\rightarrow {\bf A\underline{\otimes} A}$
satisfying the dual of the axioms of an algebra category
(coassociativity, etc.). Here, $\underline{\otimes}$ is the bicategorical
tensor product of {\bf VECT} modules (cf. [16]).

\begin{defin}
 A (not-necessarily unital or counital) {\bf bialgebra category} is a
{\bf VECT} module category
which is both an algebra and a coalgebra category, together with a
consistency natural isomorphism as in Figure 28 .

The associativity and coassociativity isomorphisms and the consistency

map $\alpha : \Delta(A) \otimes \Delta(B) \cong \Delta
(A\otimes B)$ must satisfy the following four commuting cubes given in
Figure 29 as
coherence relations.

\end{defin}

The first two of these are the Mac Lane pentagon and the dual relation
for comultiplication. The latter two first appear in [12].

Finally, a {\bf  Hopf category} is a semisimple bialgebra category
together with the
categorical analogs of a unit object, counit functor, and an antipode functor.
For brevity, we only sketch the of the
corresponding coherence relations:  the unit object is the identity
for the monoidal (algebra) category structure, and dually for the
counit functor, the counit functor is a monoidal functor, and finally
the antipode functor, satisfies the usual equations up to natural
isomorphism, with coherence conditions given by requiring that it
be a monoidal functor between the category with the given algebra
structure, and the category with its monoidal struture reversed, and
dually for the coalgebra structure.

Note, this is a slightly different (and apparently more restrictive)
notion of Hopf category than was given in [12], there the categorical
analogues of structures equivalent to an antipode in the
finite-dimensional case, but easier to construct concretely
in terms of categories were used as the definition.  We take the
{\em a priori} more restrictive notion, because our theorem for
4D TQFT's holds with this notion, and is thus {\em a priori} a
stronger result than the corresponding theorem with the definition
of [12] would be.

\section{ FACTORIZABLE 4D-TQFT AND THE HOPF CATEGORY \label{fact4}}

The definition of a factorizable 4D-TQFT is closely analogous to the
3D concept. The new element is that we can consider the gauge groups
not only on codimension 2 manifolds, i.e. on surfaces, but also on
surfaces with boundary. Thus, the category of representations of the
gauge group of the surface with boundary is acted on functorially by
the category of representations of the gauge group on the union of
circles, since there is a restriction map on gauge groups. This means
that the category of representations of the gauge group of a surface
with boundary is an element of the category of representations of the
category of representations of the gauge group on the one dimensional
boundary, which has the natural structure of a bicategory.

(We apologise that it is not practical to make this paper self
contained as regards category theory. See [16] for the discussion of
module categories over a tensor category. The reader who does not know
the definition of a bicategory can probably just ignore it on first
reading.)

With that as physical motivation, let us proceed to the
axiomatization of a factorizable 4D-TQFT.

 From one point of view one would like a clean formulation as was given
in the 3D case. Unfortunately (or fortunately, depending on one's tastes)
the notion of a monoidal tricategory has not been formalized, so we
will be obliged to describe the highest codimension assignments in
a more concrete fashion. To do this, we require:

\begin{defin} A (finitely generated)
linear bicategory is a bicategory which is
a finite product of copies of the bicategory {\bf VECT}-{\rm mod}.
We identify the indices of the product with generating objects which
are {\bf VECT} in the indexed coordinate and the trivial category in
all others. The tensor product of two linear bicategories $\cal A$ and
$\cal B$ has as generating objects ordered pairs of a generating object
from $\cal A$ and $\cal B$, and will be denoted ${\cal A} \odot {\cal B}$.
\end{defin}

\begin{defin} \label{4dfact} {\bf A 4D-TQFT with factorizability is
the following
collection of assignments:}

\begin{enumerate}
\item {\bf To each 1-manifold a linear bicategory.}
\item To each 2-dimensional cobordism a bifunctor from the linear bicategory
assigned to the source to the linear bicategory assigned to the target.
In particular {\bf to a surface with boundary we have assigned an object in
the linear bicategory associated to the boundary.} (Recall that this
object will in fact be an n-tuple of {\bf VECT} modules.) More particularly,
{\bf to every closed surface, there will be assigned a {\bf VECT} module.}
\item To each 3-dimensional cobordism with corners a binatural transformation
between the bifunctors assigned to the surfaces meeting at the corners. In
particular, {\bf to a 3-dimensional manifold whose boundary is a pair of
surfaces meeting on corners, there will be assigned a map between the
objects (in the bicategory assigned to the corners) assigned to the
surfaces. Likewise to a 3-dimensional cobordism between closed surfaces there
will be assigned a functor between the {\bf VECT} modules associated to
the surfaces, while to a 3-manifold with boundary there will be associated
an object in the {\bf VECT} module on the surface. And finally to
a 3-manifold without boundary will be assigned an object in {\bf VECT}
(i.e. a vector space).}
\item To each 4-dimensional cobordism with corners (in codimensions 2 and
3) a modification between the binatural transformations assigned
to the 3-manifolds with corners meeting at the codimension 3 corners.
In particular, {\bf to each 4-dimensional cobordism with corners
(in codimension
2) between 3-manifolds with boundary (with common boundary---the codim 2
corner) is assigned a map in the {\bf VECT} module assigned to the
corner surface between the objects assigned to the 3-manifolds with
boundary.  In the case of a 4-dimensional cobordism between closed
3-manifolds, this is a linear map between the vector spaces assigned
to the 3-manifolds.  Most particularly 4-manifolds with boundary
are assigned elements in the vector space assigned to the bounding
3-manifold and closed 4-manifolds are assigned numbers.}
\end{enumerate}

Moreover, these assignments satisfy

\begin{enumerate}
\item {\bf Disjoint unions are assigned the $\odot$ product of what is
assigned to the parts.}

\item The assignment is a trifunctor from the tricategory whose objects
are 1-manifolds, 1-arrows are 2-dimensional cobordisms, 2-arrows are
3-dimensional cobordisms with corners and 3-arrows are 4-dimensional
cobordisms with corners in codimensions 2 and 3 (with the obvious
compositions) to the tricategory whose objects are linear bicategories,
1-arrows are bifunctors, 2-arrows are binatural transformations, and
3-arrows are modifications (cf. [18]). In particular, the
assignment carries gluing along 1-dimensional strata (codimension 3
corners) of boundaries
to composition of bifunctors,  carries gluing along 2-dimensional strata
(corners)  to composition of binatural transformations, and carries
gluing along 3-dimensional strata of boundaries to composition of
modifications, and in each of the first two cases to the induced
composition on binatural transformations and modifications. More
particularly {\bf given a closed surface obtained by gluing two surfaces
with boundary, the VECT module assigned to the surface is the
hom-category in the linear bicategory assigned to the cut between
the objects therein assigned to the surfaces with boundary; given a
closed 3-manifold obtained by gluing two 3-manifolds with boundary,
the vector-space associated to the 3-manifold is the hom-space
in the VECT module assigned to the cut between the objects therein
assigned to the 3-manifolds with boundary; and given an closed 4-manifold
obtained by gluing two 4-manifolds with boundary, the number assigned
to the 4-manifold is the dual pairing between in the vector space
assigned to the cut between the vectors therein assigned to the
4-manifolds with boundary.}
\end{enumerate}
\end{defin}

Now we can  prove our main theorem for 4D-TQFTs:

\begin{thm} \label{thirdthm} In any factorizable 4D-TQFT the category
associated to the torus with a disk removed is a Hopf-category
object in the tensor bicategory associated to the circle. Moreover,
the category associated to every surface with a single boundary component
admits both an action and a coaction of this Hopf-category object
which satisfies the left-right crossed bimodule axiom up to canonoical
isomorphism.
\end{thm}

\noindent{\bf Proof:} The proof is essentially a restatement of the proofs of
Theorems \ref{firstthm} and \ref{secondthm}.
The once punctured torus now is assigned a semisimple
linear category lying in some concrete finitely generated tensor bicategory,
and the cobordisms which gave the structure maps
of the Hopf algebra before now give the corresponding structure
functors of the Hopf category. All of the natural isomorphisms in the
coherence structure for the Hopf category are given by (obvious) trivial
cobordisms, and the coherence equations for the
category are just equations between compositions of trivial cobordisms. $\Box$

\section{ ADDITIONAL STRUCTURES AND CONCLUSIONS \label{concl}}

The Hopf category we have constructed possesses some additional very
canonical structure. To begin, there are two natural 4D cobordisms
between the 3-manifold with corners representing the inclusion of the
identity object (resp. representing the counit functor) and
itself, corresponding to adding a 1- or 2- handle on a
small loop near the edge. These maps would have to satisfy some
identities which would correspond to the laws of the Kirby calculus.

Although we will not attempt to complete the procedure here, there
appears to be a natural
way to reverse the argument we are giving and reconstruct a 4D-TQFT
from a Hopf category. We would begin by giving a combinatorial
description of 4D manifolds slightly different from but related to the
Kirby calculus. A 4D manifold or cobordism can be represented as a one
parameter family of 3-manifolds equipped with Heegaard splittings. The
family changes at a discrete set of levels, where the Heegaard
splitting is either stabilized by one handle or has its surface map
changed by a simple Dehn twist. The 3D structures would be represented
by combinations of the Hopf category and its identity, while the 4D
segments would be represented by the canonical maps we have indicated.
Identities on the canonical maps would contain the information
necessary for 4D invariance. This would probably be a more elegant
construction than the one proposed in [12].

The methods outlined in this paper also could be used to construct a
4D-TQFT by a different line of attack. It was suggested in [12], and
discussed in more detail in [27], that there would be three natural
algebraic constructions of 4D-TQFT, using monoidal bicategories, Hopf
categories, and conjectural structures called ``trialgebras.''
The bicategory lives on the circle, while
the Hopf category lives on the punctured torus as we have explained in
this paper.

The trialgebra is also not hard to locate. It lives on a 3-torus from
which a solid torus with a corner has been removed. The
three algebraic operations correspond to the three obvious generators
for the homology of the 3-torus in a way analogous to the
correspondence between the two generators and the product and coproduct
(product in the dual) of the Hopf algebra associated to the 2-torus
with a disk removed.

Of course, we still obtain only a trialgebra object in a category. Even
if Witten's monopole equations allow us to construct our trialgebra
object, finding an honest trialgebra would be of comparable difficulty
to finding a quantum group from the category of representations of a
loop group at a given central extension. Motivated by the success of
that effort, we state the following conjecture:

\begin{conj} There exist a family of trialgebras from whose
representations the quantum groups can be recovered. It is also
possible to reproduce Donaldson-Floer theory from them.

\end{conj}

The existence is the hard part here, since geometric arguments of the
sort described in this paper will make the reconstructions fairly
straightforward.

\bigskip

We wish to thank Igor B. Frenkel for helpful discussions. I. Singer
and C. Taubes enlightened us as to the implications of the Witten
monopole equation.
\newpage

\centerline{\Large \bf References}
\bigskip
\bigskip

[1]  M. Atiyah
, Topological Quantum Field Theory
, Cambridge University Press
, 1990.

[2]  E. Witten
, Topological quantum field theory
, Comm. Math. Phys.
117 , 353-386, 1988.

[3] R. Dijkgraf and E. Witten, Topological Gauge Theories and Group
Cohomology, Commun. Math. Phys 129 393-429, 1990

[4]  M. Fukama, S. Hosong, and H. Kawai
, Lattice Topological Field Theory
, preprint,  1993

[5]  N. Reshetikhin and V.G. Turaev
, Invariants of 3-manifold via link polynomials and quantum
groups
, Invent. math.
103 547-597,  1991.

[6] L. Crane, 2-d Physics and 3-d Topology, Commun. Math. Phys.
135 615-640, 1991.

[7]  V. Turaev and O. Viro
, State Sum Invariants of $3$-manifolds and Quantum $6j$ symbols
, Topology
 31 865-902,  1992

[8] K. Walker, On Wittens 3 Manifold Invariants, unpublished

[9] D.N. Yetter, State-sum invariants of 3-manifolds associated to artinian
semisimple tortile categories, Top. and its App. 58 47-80, 1994.

[10] G. Kuperberg
, Involutory Hopf Algebras and Three-manifold Invariants,
 Int. J. Math. p41-61, 1989

See also : Non-Involutory Hopf algebras and 3-Manifold Invariants,
preprint, U. of Chicago math department

[11] S. Chung, M. Fukama, and A. Shapere,
, Structure of Topological Field Theories in Three Dimensions
preprint

[12] L.Crane and Igor B. Frenkel Four Dimensional Topological Quantum
Field Theory, Hopf Categories, and the Canonical Bases J. Math
Physics, special issue, p5136, 1994

[13]  G. Lusztig
, Introduction to Quantum Groups
, Birkhauser ,  1993

[14] E. Witten Monopoles and Four Manifolds, hep-th preprint 9411102

[15] C. Taubes, personal communication

[16] M. Kapranov and V. Voevodsky, Braided Monoidal 2-Categories,
2-Vector Spaces and Zamolodchikov's Tetrahedra Equation,
preprint.

See also: 2-Categories and Zamolodchikov Tetrahedra Equations, Proceedings
of Symposia in Pure Mathematics 56, Am. Math. Soc. 1994, 177-259.

[17] J. Benabou, Introduction to bicategories, in Reports of the Midwest
Category Seminar, Lecture Notes in Math 47, Springer-Verlag, 1967, 1-77.

[18] R. Gordon, A.J. Power, and R. Street, Coherence for tricategories,
preprint, 1993.

[19] D.N. Yetter, Categorical linear algebra---A setting for questions
from physics and low-dimensional topology, preprint, 1993.

[20] D. Kazhdan, lectures at Harvard University, Fall Semester 1990.

[21] R. Lawrence, Triangulations, categories and extended topological
field theories, in Quantum Topology (R.A. Baadhio and L.H. Kauffman, eds.)
World Scientific 1999, 191-208.

[22] S. Majid, Algebras and Hopf algebras in braided gategories, in
Advances in Hopf Algebras, Plenum 1993.

[23] D.N. Yetter, Quantum groups and representations of monoidal categories,
Math. Proc. Camb. Phil. Soc. 108 261-290, 1990.

[24] D. Radford and J. Towber, Yetter-Drinfel'd categories associated to
an arbitrary bialgebra, J. Pure and App. Alg. 87 259-279, 1993.

[25] S. Mac Lane, Categories for the Working Mathematician, Springer-Verlag,
1971.

[26] M. Laplaza, Coherence for distributivity, Lecture Notes in Math 281,
Springer-Verlag, 1972, 29-65.

[27] L. Crane and Igor B. Frenkel, On The Dimensional Ladder, to  appear.

\end{document}